\documentclass[aps,prc,twocolumn,superscriptaddress]{revtex4}
\usepackage{graphicx}
\usepackage{dcolumn}
\usepackage{bm}
\usepackage{amsmath}
\usepackage{epsfig}
\begin{document}

\title{Asymptotic normalization coefficients from transfer reaction and R-martix analysis of direct capture in $^{22}$Ne(p,$\gamma$)$^{23}$Na reaction}

\author{Rajkumar Santra}
\affiliation{Saha Institute of Nuclear Physics, 1/AF, Bidhan Nagar, Kolkata 700064, INDIA}
\affiliation{Homi Bhabha National Institute, Anushaktinagar, Mumbai 400094, INDIA} 
\author{Suprita Chakraborty}
\affiliation{Belgachia Aswini Dutta Vidyapith for Girls High, Belgachia Howrah, INDIA } 
\author{Subinit Roy}{\email{subinit.roy@saha.ac.in}
\affiliation{Saha Institute of Nuclear Physics, 1/AF, Bidhan Nagar, Kolkata 700064, INDIA} 

\date{\today}
 
\begin{abstract}
  
The $^{22}$Ne(p,$\gamma$)$^{23}$Na reaction in NeNa cycle plays an important role in the production of only stable sodium isotope $^{23}$Na. This nucleus is processed 
by the NeNa cycle during hot bottom burning (HBB) in asymptotic giant branch (AGB) stage of low metallicity intermediate mass stats (4 M$_O$ $\leq$ M $\leq$ 6 M$_O$).  
Recent measurements have addressed the uncertainty in the thermonuclear reaction rate of this reaction at relevant astrophysical energies through the identification
of low lying resonances at E$_p$ = 71,105, 156.2, 189.5 and 259.7 keV. In addition, precise measurements of low energy behaviour of the non-resonant capture has also 
been performed and the contribution of the sub-threshold resonance at 8664 keV excitation in $^{23}$Na has been established. Here, in this article, we have presented
a systematic R-matrix analysis of direct capture to the bound states and the decay of the sub-threshold resonance at
8664 keV to the ground state of $^{23}$Na. A finite range distorted wave Born approximation (FRDWBA) calculation has been performed for $^{22}$Ne($^3$He,d)$^{23}$Na 
transfer reaction data to extract the asymptotic normalization coeeficients (ANC-s) required to estimate the non-resonant capture cross sections or astrophysical S-factor 
values in R-matrix analysis. Simultaneous R-matrix analysis constrained with ANC-s from transfer calculation reproduced the astrophysical S-factor data over a wide energy 
window. The S$_{tot}^{DC}$(0) = 48.8$\pm$9.5 keV.b compares well with the result of Ferraro, {\it et al.} and has a lower uncertainty. The resultant thermonuclear reaction 
is slightly larger in 0.1 GK $\le$ T $\le$ 0.2 GK temperature range but otherwise in agreeent with Ferraro, {\it et al.}.
 
\end{abstract}
\maketitle

\section{Introduction}
Proton capture reaction $^{22}$Ne(p,$\gamma$)$^{23}$Na of the neon-sodium cycle of hydrogen burning in stars consumes $^{22}$Ne, a seed nucleus for neutron production
for $s$-process nucleosynthesis and converts to $^{23}$Na, the only stable isotope of sodium. The reaction occurs in the convective envelop of massive ($M \ge 4M_{O}$) 
asymptotic giant branch (AGB) stars at temperature $T \sim 1.0 \times 10^8$K ($T_9 \sim 0.1$). In more massive stars ($M \ge 50M_{O}$), the reaction takes place in 
the surface layer along with the carbon-nitrogen-oxygen (CNO) and magnesium-aluminium (Mg-Al) cycles of hydrogen burning at temperature upto $T_9 \sim 0.8$) 
\cite{cavallo, cottrell, pilachowski, paltoglou}. Since 
oxygen is destroyed in CNO cycle and Na is produced in Ne-Na cycle, the reaction $^{22}$Ne(p,$\gamma$)$^{23}$Na is said to be responsible for the observed anti-correlation 
in surface oxygen and sodium abundances in galactic globular clusters \cite{Carretta, Lind, yong}.

The rate of $^{22}$Ne(p,$\gamma$)$^{23}$Na capture reaction at relevant astrophysical energy domain is dominated by the contributions of 
several low energy resonances in $^{23}$Na and a slowly varying off-resonant capture contribution. Precision measurements 
have been carried out in recent years to identify and confirm the important low energy resonances and determine the resonance strengths \cite{cavanna1,depalo, 
cavanna2,bremmer,kelly,ferraro}. The studies helped resolve the discrepancy that existed between the recommended value of the reaction rate from the 
NACRE I compilation \cite{angulo} and those determined in Refs. \cite{Hale,iliadis1,STARLIB} in AGB stars in the temperature window 0.08 GK $\le T \le$ 0.25 GK.

The measurement of Ferraro, {\it et al.} \cite{ferraro}, besides providing the stringent upper limits for the strengths of low energy resonances at proton beam 
energies of 71 and 105 keV, also reported a slowly varying non-resonant component having significant contribution in the same temperature window. Prior to this work,
the non resonant, direct capture processes in $^{22}$Ne(p,$\gamma$) reaction to the bound states of $^{23}$Na and their contributions to low energy behaviour of 
astrophysical $S$-factor have been studied in Refs. \cite{Rolfs,gorres1,kelly}. Earlier studies used a constant value of $S_{DC}$(E) = 62.0 keV.b obtained by 
G$\ddot o$rres, {\it et al.} \cite{gorres1} from their measurement of higher energy direct capture data. Kelly, {\it et al.} \cite{kelly} measured the direct 
capture cross section at a single proton beam energy of 425 keV and maintained the same value for $S_{DC}$(E=0 MeV). Ferraro, {\it et al.} in a subsequent high 
statistics and low background measurement extended the non-resonant capture cross section to 310, 250, 205 and 188 keV proton beam energies. Their analysis 
yielded $S_{DC}$(0) = 50 $\pm$ 12 keV.b. However, the authors observed a distinct rise in the low energy astrophysical $S$-factor data and showed it to  
in the low energy non-resonant astrophysical $S$-factor is shown to be the consequence of capture to the sub-threshold state at 8664 keV excitation in $^{23}$Na.
Thus the direct capture component along with the contribution of sub-threshold state actually determines the total off-resonant component around the temperature of
$T_9 \sim 0.1$.

In the present article, we reported a detailed analysis of the available $S$-factor data for non-resonance capture to the ground state and 440, 2392, 2982, 6318, 6918
and 8664 keV excited states of $^{23}$Na nucleus and the total non-resonant $S$-factor data within the framework of $R$-matrix model. The aim is to investigate the 
energy dependence of the total off-resonance astrophysical $S$-factor including the contribution of broad, sub-threshold (-130 keV) state at 8664 keV 
and to extend the curve to still lower energy of astrophysical interest. The spectroscopic information required for the $R$-matrix calculation for the direct capture
contribution have been extracted from the re-analysis of one proton 
transfer reaction data on $^{22}$Ne from the literature. The total reaction rate is then estimated from the calculated reaction rate due to off-resonance process and 
using the measured strengths of important low energy resonances in $^{23}$Na except for the state with resonance energy of 151 keV. Experimental evidence shows that 
the state at 8945 MeV excitation corresponding to resonance energy 151 keV is actually a doublet with capture of $d$- and $f$- waves respectively. A reanalysis of 
proton transfer reaction to 8945 MeV unbound state ($Q_p$ = 8794 MeV) has been performed to extract the resonance strengths indirectly. 
The resultant reaction rate up to $T$ = 1.0 GK has been compared with the recent estimations.

\begin{figure}
\includegraphics[scale=0.85]{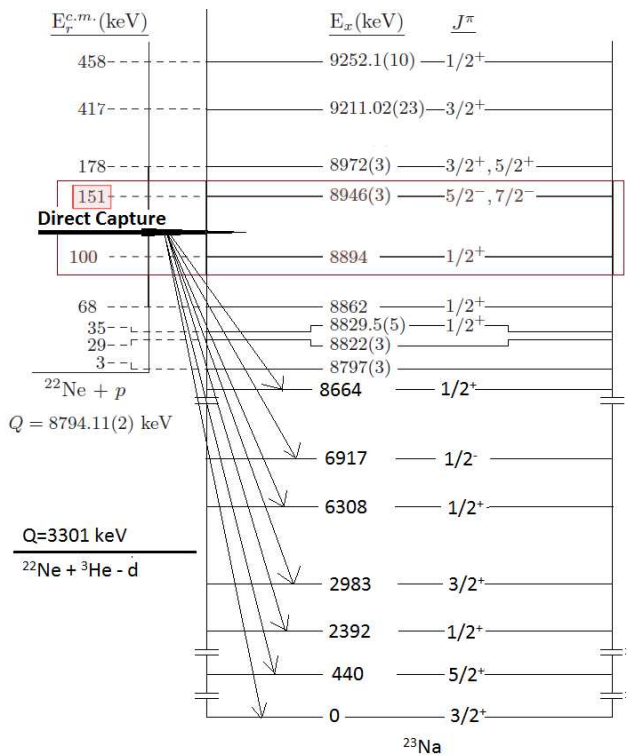}
\caption{\label{fig1} Level scheme of ${}^{23}$Na }
\end{figure}

\begin{table*}
\caption{Presnt status of $^{22}Ne(p, \gamma)^{23}Na$ reaction rate.}
\label{tab1}
\begin{tabular}{ccccccccc}    \\ \hline \hline
Year&&& S$_{DC}$(0)&&&&Reaction rate(T$_9$$\sim$0.1) \\ 
&&&&&&&& \\  
&&&(keV.b)&&&&cm$^3$mol$^{-1}$s$^{-1}$ \\ \hline
&&&&&&&& \\
Rolfs 1975 \cite{Rolfs}&&&67$\pm$19&&&&- \\ 
&&&&&&&& \\
Gorres 1983 \cite {gorres1} &&&62&&&&- \\ 
&&&&&&&& \\
Hale 2001 \cite{Hale}&&&-&&&&2.1$\times$10$^{-9}$ \\ 
&&&&&&&& \\
Sallaska 2013 \cite{STARLIB}&&&-&&&&5.52$\times$10$^{-9}$\\ 
&&&&&&&& \\
Cavanna 2015 \cite{cavanna1}&&&-&&&&6.6$\times$10$^{-8}$ \\ 
&&&&&&&& \\
Depalo 2016 \cite{depalo}&&&-&&&&2.7$\times$10$^{-8}$ \\ 
&&&&&&&& \\  
Kelly 2017 \cite{kelly}&&&-&&&&3.53$\times$10$^{-8}$ \\ 
&&&&&&&& \\  
Ferraro 2018 \cite{ferraro}&&&50$\pm12$\cite{ferraro}&&&&2.1$\times$10$^{-8}$ \\ 
&&&&&&&& \\ \hline
&&&&&&&& \\  
\end{tabular}
\end{table*}

\begin{table}
\caption{Spectroscopic factors and  asymptotic normalization coefficients (ANC-s) for the first seven states of $^{23}$Na}
\label{tab2}
\begin{tabular}{ccccccc}\hline \hline    \\ \hline
E$_x$ & J$^\pi$ & nl$_j$ &C$^2$S\footnote{Geometry parameters of b.s. potential $a_0$=0.6, $r_0$=1.26} &C$^2$S &b& ANC  \\ 
(keV)&&&Present&Ref. \cite{Power}&(fm$^{-1/2}$)&(fm$^{-1/2}$)  \\ \hline
&&&&& \\
g.s. & 3/2$^+$ & 1d$_{3/2}$&0.082$\pm$0.012&0.08&6.86& 1.96$\pm$0.5  \\
&&&&&  \\
440  &5/2$^+$  & 1d$_{5/2}$&0.38$\pm$0.08&0.35&7.62 &4.69$\pm$0.8 \\
&&&&& \\
2392 &1/2$^+$  & 2s$_{1/2}$ &0.26$\pm$0.05&0.25&17.56& 8.8$\pm$1.6  \\
&&&&& \\
2982 &3/2$^+$  & 1d$_{3/2}$ &0.35$\pm$0.04&0.32&4.38 &2.59$\pm$0.87   \\
&&&&& \\
6308 &1/2$^+$  & 2s$_{1/2}$&0.14$\pm$0.02&0.13&11.14&4.16$\pm$0.79   \\
&&&&& \\
6917&1/2$^-$  & 2p$_{1/2}$&0.18$\pm$0.04&0.15&7.27 &3.1$\pm$0.7   \\\hline
&&&&& \\
\end{tabular}
\end{table}

\section{Analysis}
The model calculation progressed in two steps. In the first step, the data from $^{22}$Ne($^3$He,d)$^{23}$Na transfer reaction to the bound states of $^{23}$Na
have been reanalyzed to extract the Asymptotic Normalization Coefficients or ANCs of the states. The second part constitutes of R-matrix calculation for 
the data of direct capture to those bound states using the ANCs from the first part of the calculation.

\subsection{Finite Range DWBA analysis and extraction of ANC}

A Finite Range Distorted Wave Born Approximation (FRDWBA) calculation has been performed for the angular distribution data of $^{22}$Ne($^3$He, d)$^{23}$Na 
one proton stripping reaction from Refs. \cite{Hale, Power}. In FRDWBA model, conventionally the experimental cross section of a transfer reaction
$A + a (=b+x) \rightarrow B (=A+x) + b$ (where $x$ is the transfered particle) is compared with the calculated cross section by the relation 

\begin{equation}
\Big(\frac{d\sigma}{d\Omega}\Big)_{exp} = (C^2S)_{bx} (C^2S)_{Ax} \Big(\frac{d\sigma} {d\Omega}\Big)_{mod} 
\end{equation}

where $C^2S_{bx}$ is the product of spectroscopic factor $S_{bx}$ and isospin Clebsh Gordon coeeficient $C^2_{bx}$ of $b+x$ configuration in projectile $a$ 
and $C^2S_{Ax}$ is that for $A+x$ configuration in residual nucleus $B$. $\Big(\frac{d\sigma} {d\Omega}\Big)_{mod}$ is the cross section obtained from model 
calculation.  

To extract the spectroscopic factors from data, a finite range distorted wave Born approximation (FRDWBA), using the code FRESCO (ver. 2.9) \cite{FRESCO}, 
has been performed. Angular distribution data, measured at 15 MeV incident energy, for transfer to ground state and 440, 2392, 2982, 6308, 6917 keV exited states 
are taken from Ref. \cite {Power}. The data from Ref. \cite{Hale}, measured with 20 MeV $^3$He beam, have been used for transfer to the sub-threshold 
state of $^{23}$Na at 8664 keV excitation energy.

In the reanalysis of 15 MeV data within the FRDWBA framework, the optical model potential parameters for entrance and exit channels are taken from 
Ref. \cite{Power}. Standard Woods-Saxon form has been used for the potentials. The shape parameters for the bound sate potentials related to $^{22}$Ne+p 
and d+p systems are from Refs. \cite{Power} and \cite{bem},
respectively. The strengths of the bound state potentials are varied to get the binding energies of the states of the composite nuclei.

The model calculations reproduced the angular distribuitons for transfer to ground state and 440, 2392, 2982, 6308, 6917 keV exited states of $^{23}$Na 
quite well. While extracting the spectroscopic factors of the states of $^{23}$Na, the spectroscopic factor $C^2S_{dp}$ for $^{3}$He is taken as 1.092, a 
value that is derived for the $^3$He ground state with $d+p$ configuration using the method reported in Ref.\cite {bem}. The resultant spectroscopic 
factors of $^{23}$Na states are shown in Table \ref{tab2}. The values obtained from the present FRDWBA analysis match well with those reported from 
zero range DWBA calculation in Ref. \cite{Power}.

\subsubsection{DWBA analysis of ${}^{22}$Ne(${}^3$He,d) reaction for E$_{x}$=8664 keV state}
The excited state 8664 keV of $^{23}$Na is 130 keV below the proton threshold at 8794 keV. Capture through this sub-threshold resonance
controls the low energy behaviour of the astrophysical S-facor of $^{22}$Ne(p,$\gamma$) reaction. To extract the spectroscopic factor of this 
state we again performed a FRDWBA calculation for the transfer reaction $^{22}$Ne($^{3}$He,d) at 20 MeV with the data from Ref. \cite{Hale}.  
Potential parameters used to obtain the transfer angular distribution are given in Table \ref{tab3}. It has been observed that unlike the more 
deeply bound states in 
$^{23}$Na, a complex remnant term is required to obtain a very good overall fit to the angular distribution data. The parameters of d+$^{22}$Ne 
core-core potential are also given in Table \ref{tab3}. The resultant fit is shown by solid red 
line in Fig. \ref{fig3}. The blue dashed dotted line represents the FRDWBA calculation without the remnant term. Improvement in the fit is quite 
remakable. In Col.5 of Table \ref{tab4}, the extracted spectrosopic factors have been shown. The value obtained from the present work is very close 
to the values reported in Refs. \cite{Hale,gorres1}.

\begin{table*}[]
\centering
\caption{Potential parameters for ${}^{22}$Ne(${}^3$He,d)$^{23}$Na (E$^*$= 8664 keV), $E_{lab}$ = 20 MeV \cite {Hale}.}
\label{tab3}
\begin{tabular}{lllllllllllll}    \\ \hline
channel     &  V$_r$ & r$_r$&  a$_r$ & W$_i$  &  W$_D$ & r$_i$=r$_D$ & a$_i$=a$_D$ & V$_{so}$ &  r$_{so}$ & a$_{so}$ & r$_c$  \\ 
	    &  (MeV) & (fm) &   (fm) & (MeV)  &  (MeV) &  (fm)       & (fm)        & (MeV)    &   (fm)    &  (fm)    &  (fm)   \\ \hline
	    &        &	    &	     &	      &	       &	     &	           &	      &	          &	     &	       \\
$^3$He+$^{22}$Ne& Ref. \cite{Hale}	    &        &	    &	     &	      &	       &	     &	           &	      &	          &	     &	       \\
d+$^{23}$Na & Ref. \cite{Hale}&  &   &        &        &             &             &          &           &          &          \\ 
d+$^{22}$Ne &  88.0  &  1.17&  0.73  &  0.24  &  35.8  & 1.33        &     0.73    & 13.85    &    1.07   &   0.66   & 1.33   \\ 
d+p         &     *  &  1.25&  0.65  &        &        &             &             &  6.2     &    1.25   &   0.65   & 1.30   \\ 
p+$^{22}$Ne & Ref. \cite {Hale}&  &  &        &        &             &             &          &           &          &         \\\hline

* Varied to match separation energy.
\end{tabular}
\end{table*}

\begin{table}
\caption{Spectroscopic factor and  asymptotic normalization coefficient (ANC) of 8664 keV state of $^{23}$Na}
\label{tab4}
\begin{tabular}{ccccccc}\hline \hline    \\ \hline
E$_x$ & J$^\pi$ & l$_p$ &nl$_j$ &C$^2$S & b & ANC  \\ 
(MeV)&&&&&(fm$^{1/2}$)&(fm$^{1/2}$)  \\ \hline
&&&&&& \\
8664\footnote{Subthreshold state. E$_{r}$=-130 keV}&1/2$^+$&0&2s$_{1/2}$&0.32$\pm$0.05&252\footnote{Geometry parameters of b.s. potential $a_0$=0.69, $r_0$=1.17}&143.7$\pm$15.2  \\
&&&&&&  \\ 
  &  & &  &0.29\cite{Hale}& &  \\
&&&&&&  \\ 
& &  &  &0.3\cite{gorres1}& &  \\
&&&&&&  \\ 
  &  &  &  &0.42$\pm$0.08\cite{ferraro}& \\
&&&&&&  \\ 
  &  &  &  &0.58$\pm$0.08\cite{Terakawa}& &  \\
&&&&&&  \\ \hline  
\end{tabular}
\end{table}

\subsubsection{Extraction of ANC}

The spectroscopic factor so determined includes the effect of nuclear interior and measures the many-body effect in the transfer reaction process.
It depends on the choice of the potentials, more sensitively on the geometry parameters of bound state potential used to describe a 
particular configuration.  
In low energy radiative capture reactions, instead of spectroscopic factor, asymptotic normalization coefficient or ANC is more relevant a quantity. 
ANC measures the amplitude of the tail of the overlap between the bound state wavefunctions of initial and final nuclei. It is related to the spectroscopic 
factor of the two-body configuration as
\begin{equation}
C^2S_{J_f l_f} = \big(\frac{C_{J_f l_f}}{b_{l_f j_f}}\big)^2
\end{equation}
where $C^2S_{J_f l_f}$ is the spectroscopic factor of the configuration in the composite nucleus with total spin $J_f$. The relative orbital angular 
momentum and spin of the two clusters in the final bound state are denoted by $l_f$ and $j_f$.  $C_{J_f l_f}$ is the corresponding ANC and 
$b_{l_f j_f}$ is the single particle asymptotic normalization constant (SPANC) with $l_f$ and $j_f$ quantum numbers of the bound state orbital used 
in the DWBA calculation. 
The SPANC 'b' is expressed in terms of bound state wave function of the composite nucleus \cite{Thomson book} as
\begin{equation}
b(r_0,a_0) =\frac {u(r,r_0,a_0)}{W_{-n,l+1/2}(2\kappa r)}   
\end{equation}
in the asymptotic radial region. W$_{-n,l+1/2}(2\kappa r)$ is the Whitteker function, $\kappa = \sqrt{2 \mu \epsilon}$ the wave number, and $\mu,\epsilon, R_N $ 
are the reduced mass, binding energy and the nuclear interaction radius, respectively for the bound state of the final nucleus. Both bound state wavefunction and 
Whitteker function have similar radial fall of in the asymptotic region. The parameters $r_0$ and $a_0$ are the radius
parameter and diffuseness of Woods Saxon potential generating the required bound state wavefunction. In all cases the strength of the 
bound state potential is obtained by reproducing the binding energy of the state. 

\begin{figure*}
\includegraphics[scale=0.5]{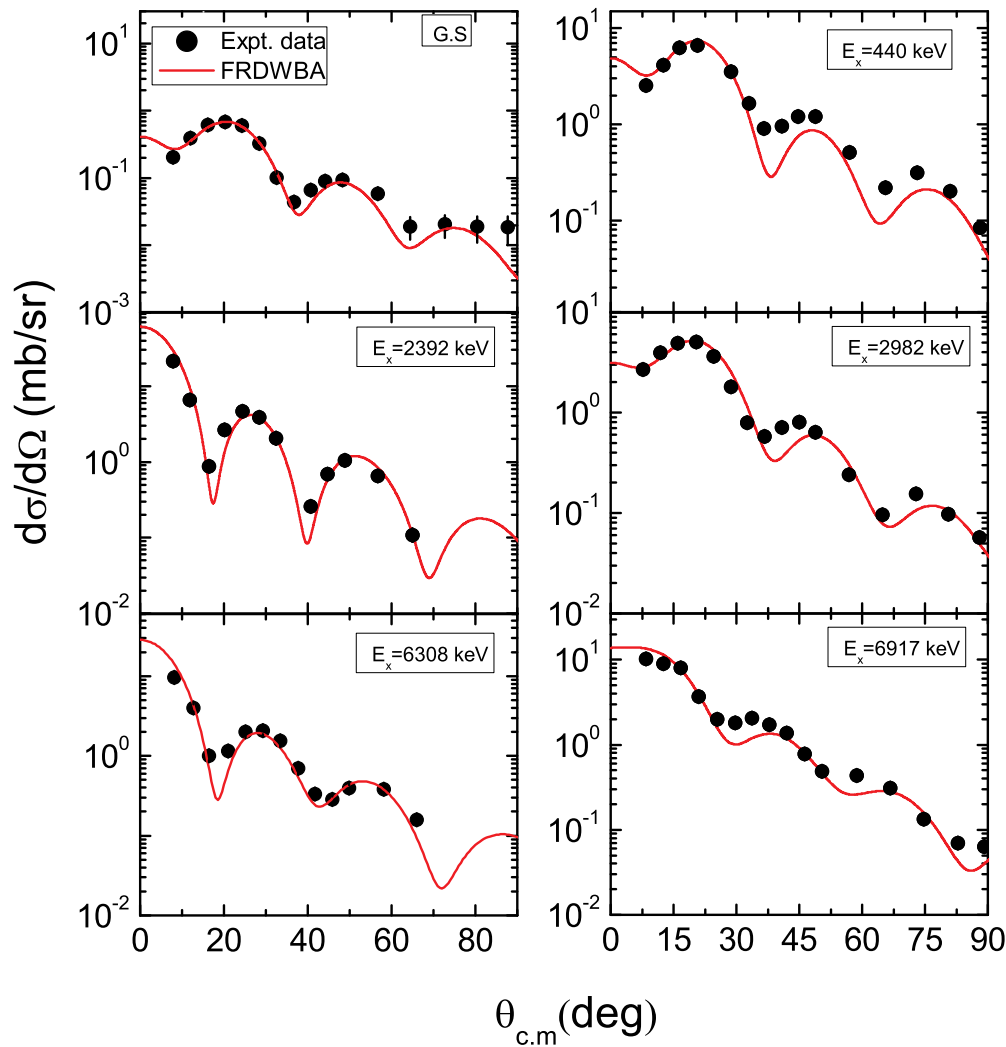}
\caption{\label{fig2}Transfer angular distributions fitted with FRDWBA model calculation.}
\end{figure*}

\begin{figure}
\includegraphics[scale=0.3]{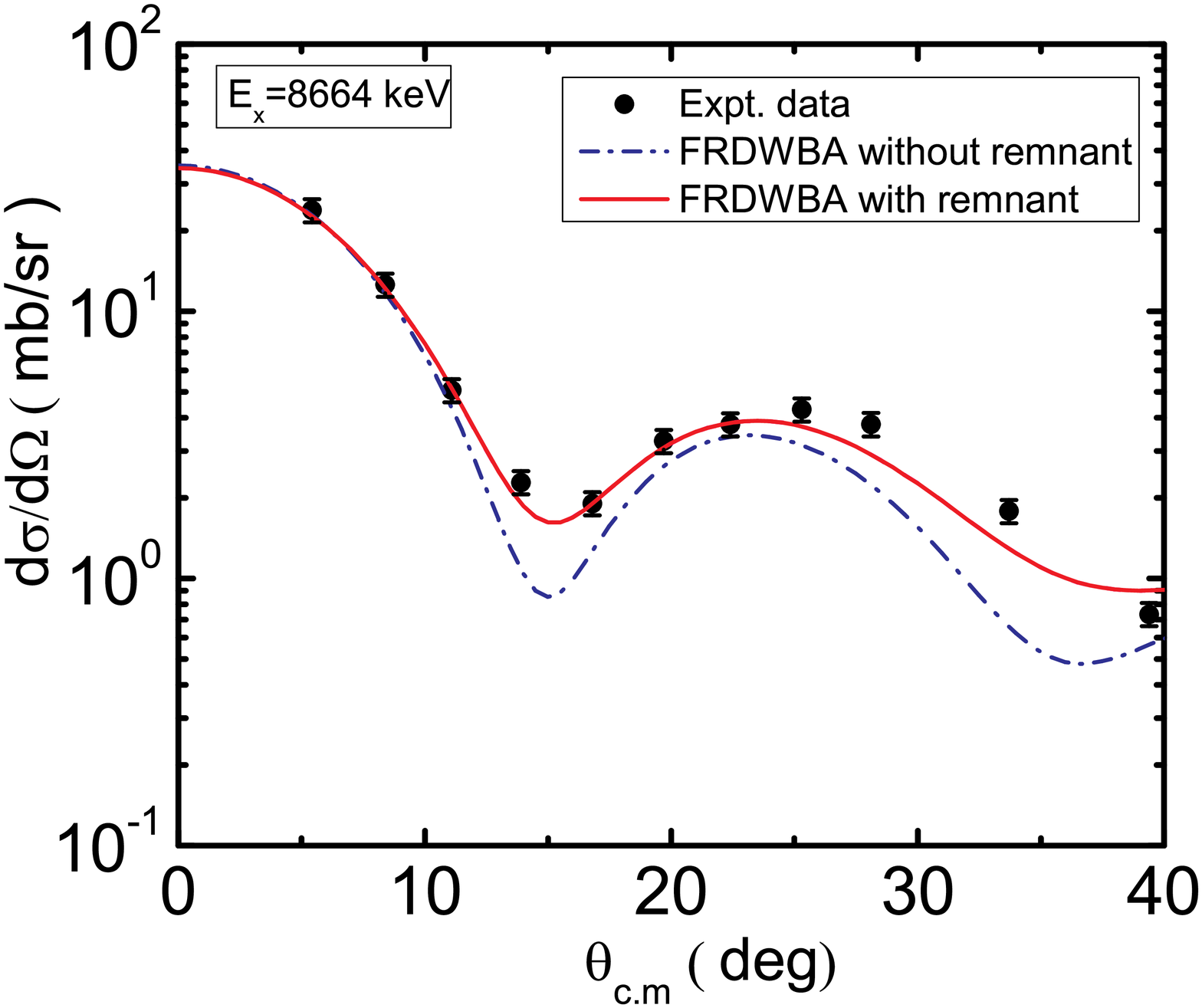}
\caption{\label{fig3} DWBA fit to angular distribution data of Power \textit{et al.} \cite{Power} for the state E$_{x}$=8664 keV. The  
line represents calculated cross section}
\end{figure}


\begin{figure}
\includegraphics[scale=0.3]{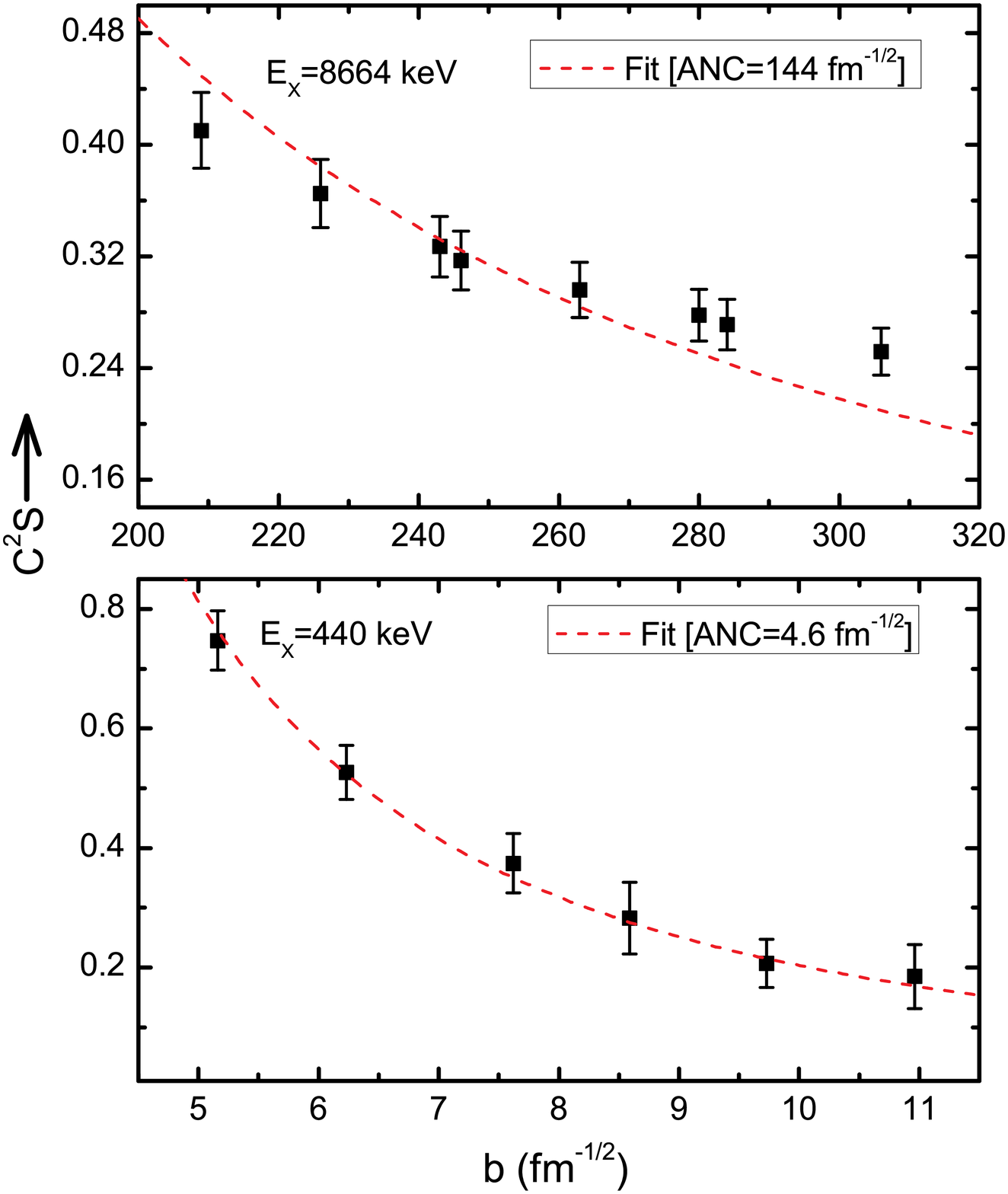}
\caption{\label{fig4} Variation of spectroscopic factor with single particle ANC for 8664 MeV (top panel) and 440 MeV (lower panel) states.} 
\end{figure}

\begin{figure}
\includegraphics[scale=0.31]{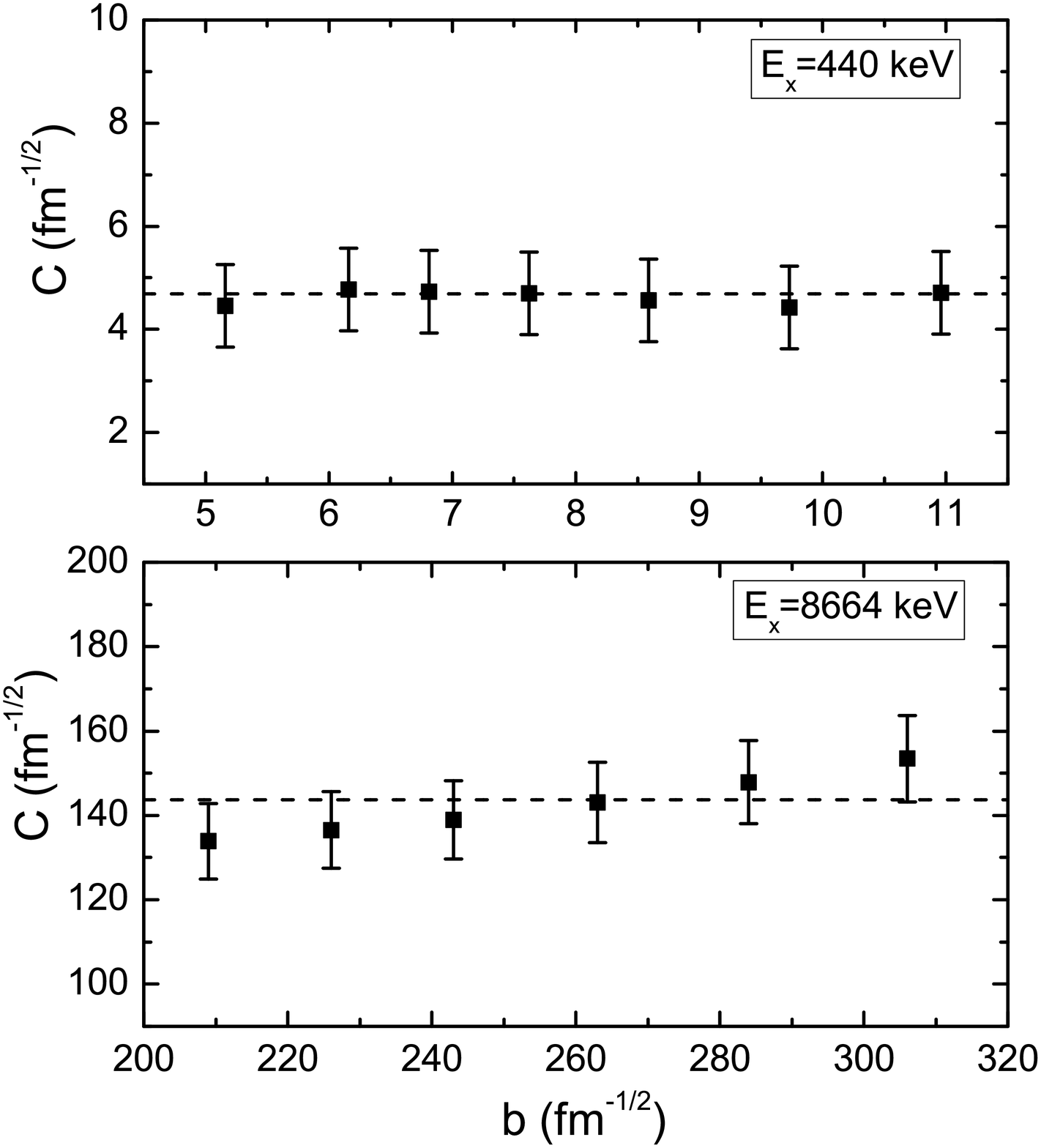}
\caption{\label{fig5}Variation of ANC (C $fm^{-1/2}$) as a function of SPANC (b $fm^{-1/2}$) for 440 keV (top panel) and 8664 keV (bottom panel) states.}
\end{figure}

\begin{figure}
\includegraphics[scale=0.3]{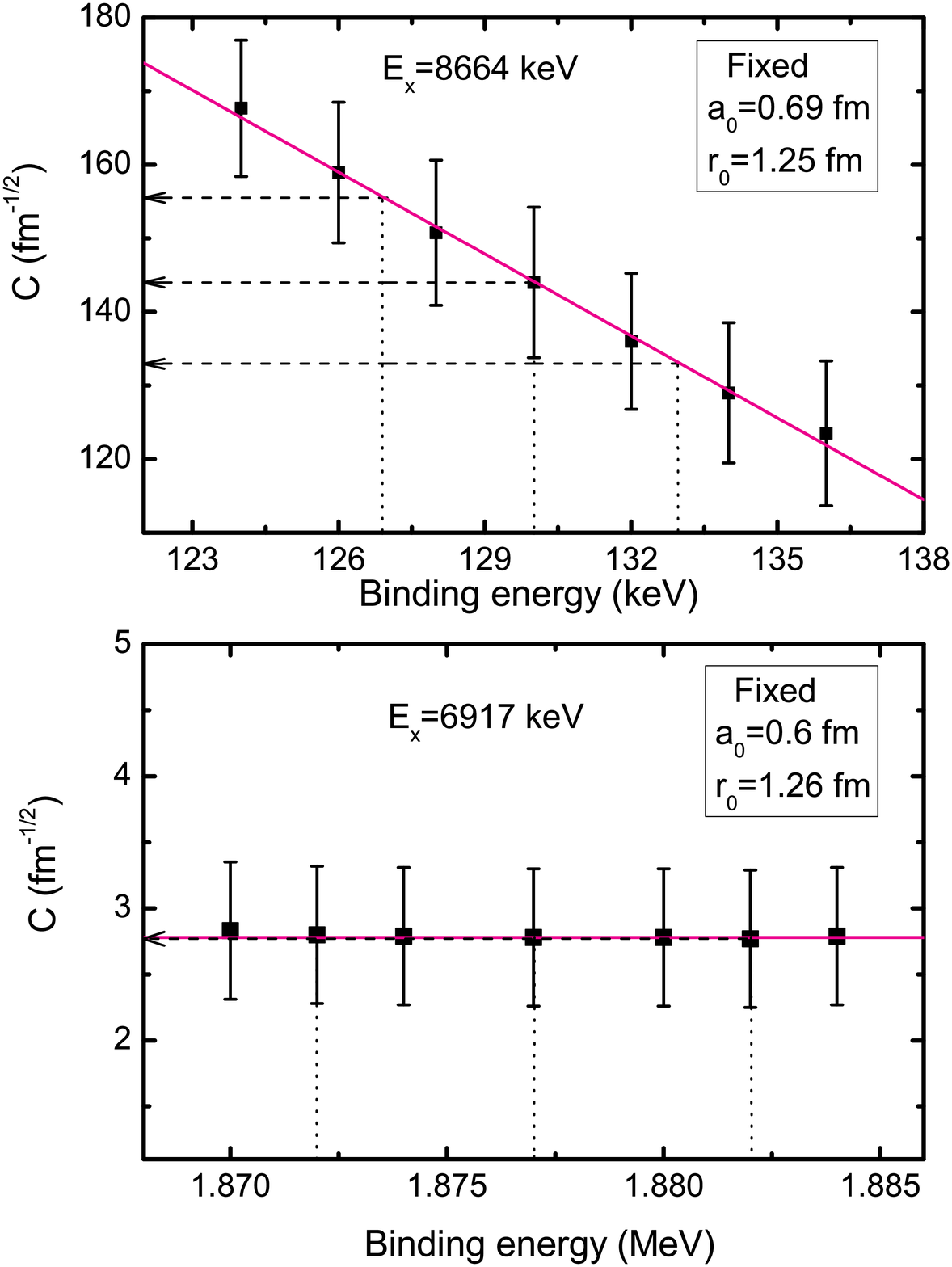}
\caption{\label{fig6} Variation of ANC (C $fm^{-1/2}$) as a function of binding energy for 8664 keV (top panel) and 6917 keV (bottom panel) 
states for fixed geometry parameters of bound state potential.}
\end{figure}

\begin{figure}
\includegraphics[scale=0.31]{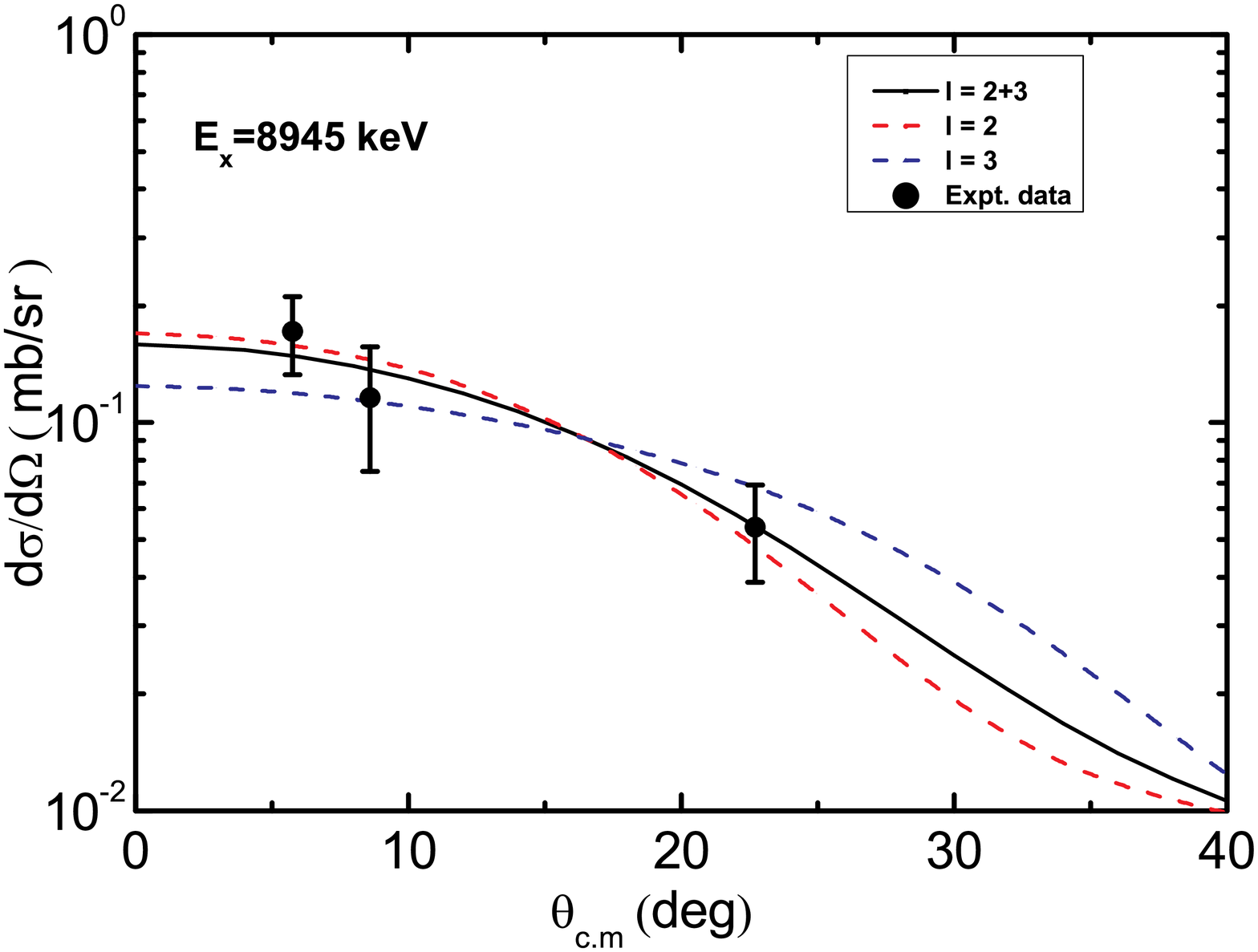}
\caption{\label{fig7} Angular distribution data and DWBA fits to 151 keV resonance state of ${23}$Na.}
\end{figure}
In the present work, the  value $b_{l_f j_f}$ associated with each bound state has been obtained from the best fit ratio value using Eqn.3 for the region beyond 
the radius R$_N$ = 5.5 fm.  
Like the spectroscopic factor, SPANC {\it b} also depends on the choice of potential parameters. But the variations of the two quantities with the geometry parameters 
of the bound state potential are opposite in nature. Hence, the product of these two quantities that gives the required ANC, remains constant with the change of 
potential parameters for a peripheral reaction. Thus, for a pure peripheral condition the variation of spectroscopic factor should be proportional to the inverse 
square of SPANC value \cite{pang} from Eq.3. In Fig.~\ref{fig4} we have shown the plots of variation of C$^{2}$S as a function of SPANC {\it b} for the 8664 keV subthreshold 
state and deeply bound 440 keV state. The error shown in the figure for C$^2$S includes the uncertainty of DWBA fit to the angular distribution data for a particular 
SPANC value obtained for the chosen r$_0$ and a$_0$ parameters of the bound state potential and the experimental error of individual cross section data.
The radius and diffuseness parameters have been changed in small steps to generate the corresponding  SPANC value. The fits with inverse square function to 
the extracted data ensures the pheripheral nature of the process and hence the correctness of the extraction of 
ANC value. In case of the state 8664 keV, which is a subthreshold state, the reproduction of the variation does not follow a purely inverse square dependence. 
The mismatch is a result of non-reproduction of the tail part of this weakly bound state with the asymptotic radial behaviour of the Whitteker 
function. The values of SPANC-s and the corresponding ANC-s of the states of $^{23}$Na have been listed in Tables \ref{tab2} and \ref{tab4}.

\subsubsection{Uncertianties of extracted ANC values}

The uncertainty of the estimated value of ANC has been calculated by propagating the error of the spectroscopic factor through the relation given in Eq.2. In Fig. \ref{fig5},
extracted ANC with the estimated error has been shown as a function of SPANC {\it b}. The fit to these secondary data points produce the mean ANC value along with its
uncertainty. Besides, the dependence of the extracted ANC on the binding energy of the state has also been checked. In Fig. \ref{fig6}, we have shown the plots of
ANC as a function of binding energy \cite {keeley} for the states at 8664 keV and 6917 keV excitation energies of $^{23}$Na. Binding energy of a state is varied
keeping the geometry parameters of the bound state potential corresponding to the mean ANC value for the state fixed. The plots show that unlike the 
more bound 6917 keV state, the ANC value of sub-threshold state 8664 keV decreases with the increasing binding energy. Thus for the 8664 keV state the uncertainty
of the ANC due to the $\pm$3 keV \cite{Hale} uncertainty in the binding energy has been estimated graphically from the plot shown in Fig. \ref{fig6}. Uncertainties 
in the ANC-s for other deeply bound states corresponding to the error in the binding energies are negligibly small and not considered.

\begin{table*}
\caption{Angular momentum transfers, spectroscopic factors and proton widths of 8945 MeV state in $^{22}$Ne($^3$He,d)$^{23}$Na reaction.}
\label{tab5}
\begin{tabular}{ccccccc}    \\ \hline
E$_x$ & J$^\pi$ & l$_p$ & nl$_j$ & C$^2$S &C$^2$S (Litt.)& $\Gamma_p$ (keV) \\ 
(keV)&&&&Present&Ref.\cite{Hale}& \\ 
&&&&&& \\ \hline
8944 & 3/2$^+$ & 2 & 1d$_{3/2}$ &(5.54$\pm$1.41)$\times$10$^{-4}$ & 8.32$\times$10$^{-4}$&(9.99$\pm$2.50)$\times$10$^{-8}$\\
&&&&&& \\                             
8945 & 7/2$^-$ & 3 & 1f$_{7/2}$ & (3.94$\pm$0.9)$\times$10$^{-4}$& $\leq$1.08$\times$10$^{-3}$&(9.83$\pm$2.24)$\times$10$^{-10}$  \\                              
\hline  
\end{tabular}
\end{table*}

\subsubsection{DWBA calculation for E$_x$=8945 keV resonance state}

The state at excitation of 8945 keV in $^{23}$Na is a resonance state about 151 keV above the proton threshold. It has an important contribution 
in the reaction rate of $^{22}$Ne(p, $\gamma$) reaction at T = 0.1 GK as it falls within the Gamow window at this temperature. In earlier reports
\cite{gorres,Hale}, it was considered that at this excitation a single state exists with generally adopted spin parity of 7/2$^-$. Later Jenkins, 
et al.\cite{Jenkins} in their $\gamma$ spectroscopic study of $^{23}$Na have shown that at this excitation the nucleus has a doublet of states with 
about a keV difference in excitation energy. One of them has a spin parity of $J^\pi$ = 7/2$^-$ and decays to 9/2$^+$ and 5/2$^+$ states of $^{23}$Na 
by dipole transitions. The assignment is consistent with a $l$ = 3 angular momentum transfer from a ($d,n$) study \cite{Childs}. On the otherhand, the 
measurement also shows a distinct coincidence of a 3914 keV $\gamma$ ray depopulating the 5/2$^+$ state at 3914 keV excitation with a 5030 keV $\gamma$ ray 
that depopulates the relevant 8944 keV state. A spin parity of 3/2$^+$ has been assigned to this second state from its decay branches and angular correlation 
ratio. In a further study, Kelly, {\it et al.} \cite{kelly} have also observed a strong primary transition from this 3/2$^+$ state to the 5/2$^+$ 3914-keV state 
with branching ratio of 80$\%$ and to the 1/2$^+$ 2391 keV state with 20$\%$. Authors have also performed zero-range DWBA fits to the data of 
$^{22}$Ne($^3$He, $d$)$^{23}$Na$^*$ (8945 MeV) reaction \cite{Hale} with $l$=1,2,3 angular momentum transfers. They opted for $l$=2 transfer assigning 
a 3/2$^+$ spin-parity for the 8945 MeV state.

Although the number of data points in the angular distribution is small, we carried out a re-analysis within the zero range DWBA framework for this unbound state using 
the code DWUCK4 code \cite{DWUCK4}. Same set of potenials from Ref. \cite{Hale} is used. Values obtained by Hale, {\it et al.} assuming a $l$=3 transfer and by Kelly, {\it et al.} \cite{kelly} assuming $l$=2 are reproduced. Subsequently, we completed a least square fit to the angular distribution data assuming that both $l$=2 ($J^\pi$=3/2$^+$) and $l$=3 ($J^\pi$=7/2$^-$) can contribute and the calulation yielded the spectroscopic factors shown in Col.5 of Table \ref{tab5}. The fits obtained are compared in Fig. \ref{fig7}. A normalizing constant N=4.42 has been used in the zero range DWBA calculation for ($^3$He,d) \cite{Hale, Power}. 
It is apparent from Fig. \ref{fig7} that the fit obtained considering the contributions of both $l$=2 and 3 is a better reproduction of the limited angular distribution data available. 

The partial widths $\Gamma_{p}$ have been estimated for the doublet states having excitation energy E$_{x}$= 8945 MeV from the extracted spectroscopic factors using
the relation
\begin{equation}
\Gamma _p = (C^2S) \Gamma _{sp}
\end{equation}
where C$^2$S) is the spectroscopic factor of the resonant state of ${}^{23}$Na for the particular configuration and  $\Gamma _{sp}$ is single particle 
width of the state. Single particle width $\Gamma _{sp}$ for a pure single particle configuration depends, like the spectroscopic factor, on the choice 
of the nuclear potential used to generate the corresponding wavefunction. To estimate the systematic uncertainty in extracted $\Gamma_p$, we varied the 
radius and diffuseness parameters of the bound state potential from 1.125 to 1.375 fm and from 0.39 to 0.89 fm, respectively keeping the binding energy 
fixed. It is observed that partial width $\Gamma_p$ is more-or-less independent of the chosen parameters as C$^2$S and $\Gamma_{sp}$ have opposing trends 
of dependence on the parameters. In the last column of Table \ref{tab5}, the extracted particle widths have been listed. The error shown include the 
fitting uncertainty as well as the systematic uncertainty. We retained the individual contributions of the doublet pair in the estimation of rate of 
proton capture reaction within the relevant temperature window. 

\begin{table*}
\caption{Background pole parameters obtained from R-matrix fits}
\label{tab6}
\begin{center}
\begin{tabular}{cccccccccccc}
\hline \hline 
J$^\pi$&E$_x$ &$\Gamma _P$& \multicolumn{9}{c} {$\Gamma_\gamma$[E1]} \\ 
&(MeV) &(MeV)& \multicolumn{9}{c} {(eV)} \\  \hline
&&& R$\rightarrow$g.s &R$\rightarrow$0.44&R$\rightarrow$2.39 &R$\rightarrow$2.98&R$\rightarrow$6.30 &R$\rightarrow$6.91 &R$\rightarrow$8.66  \\
&&&&&&&&&&& \\  \hline  \hline 
&&&&&&&&&&& \\ 
1/2$^-$&15&5.0 & 589.92 &- & 2.77$\times$10$^{3}$ & 499.10 & 912.64 &-&118.42 \\
&&&&&&&&&&& \\       
1/2$^+$&15&5.0 &-&-&-& -& -& 4.41 &  \\ 
&&&&&&&&&&& \\    
3/2$^-$&15&5.0 &-& 632.97& -&-&- &-\\ 
&&&&&&&&&&& \\ \hline
\end{tabular}
\end{center}
\end{table*}

\begin{figure*}
\includegraphics[scale=0.6]{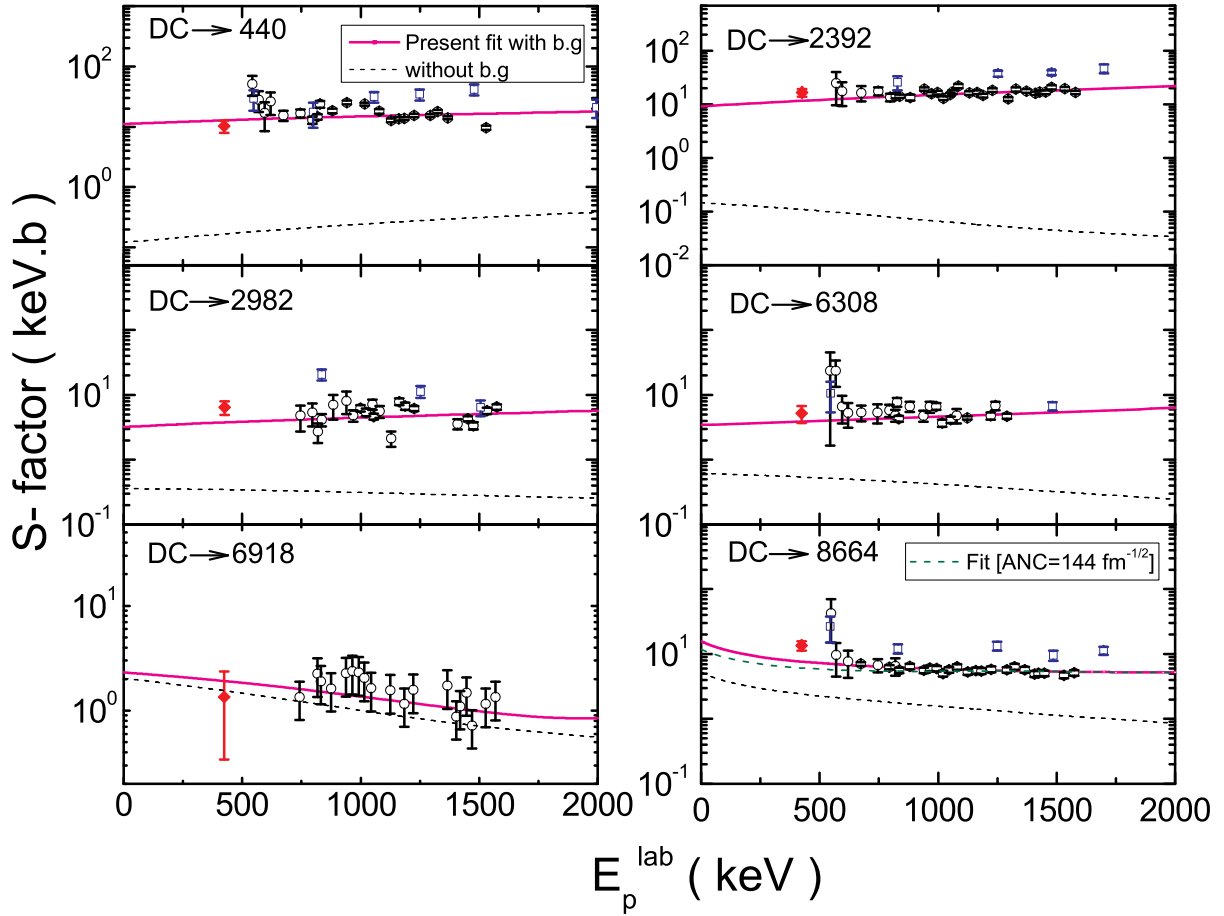}
\caption{\label{fig8} R-matrix fit to the S-factor curve for direct capture to six bound states of {}$^{23}$Na. The red solid curves are 
the R-matrix fits with the contribution of background poles while the dashed curves represent the caculation without the background poles. 
The panel showing the S-factor curve for DC$\rightarrow$8664 keV sub-threshold state also includes the R-matrix fit (green dashed) with ANC 
value fixed from transfer reaction calculation. The solid curve in this panel depicts the R-matrix fit with ANC value of 166 fm$^{-1/2}$ for the state.}
\end{figure*}

\subsection{R matrix calculation for direct capture process}

The low energy behaviour of off-resonance astrophysical S-factor for $^{22}Ne(p, \gamma)^{23}Na$ reaction is determined by the direct capture 
process and a broad sub-threshold resonance at 8664 keV in compound nucleus $^{23}$Na \cite{ferraro}. Present work attempts a R-matrix description of the low
energy behaviour of the off-resonance S-factor through the estimation of direct capture component and the contribution of the sub-threshold state
constrained by the extracted asymptotic normalization coefficients (ANC-s).   

The modeling of direct capture in $^{22}Ne(p, \gamma)^{23}Na$ is done using the R-matrix code AZURE2 \cite{azure} based on the basic theory 
developed in the seminal works of Lane and Thomas \cite{lane-thomas} and of Vogt \cite{vogt}.
In R-matrix modeling, the {\it channel radius} ($r_c$) divides the radial space into an external and an internal parts \cite{lane-thomas}. 
Accordingly, the capture cross section is divided into an external capture contribution coming from the radial region beyond $r_c$ and an internal 
capture contribution from the region below $r_c$. 
The magnitude of the external capture cross section is determined by the {\it asymptotic normalization coefficient} (ANC) of the final bound 
state \cite{azure, angulo, Kontos}. The internal capture component of the direct or
non-resonant contribution, on the other hand, is simulated by the high energy background states in the composite nucleus. Thus the direct capture part 
of the cross section is modelled as a sum of external capture component and the contribution from high energy background poles in AZURE2. 

In the present work, we have fitted simultaneously the direct capture data of Rolfs \textit{et al.} and G$\ddot{o}$ress \textit{et al.}. 
A detailed experimental study of nonresonant or direct capture component of the reaction $^{22}Ne(p, \gamma)^{23}Na$ by Rolfs \textit{et al.} \cite{Rolfs} 
reported the measurement of cross sections for the transitions to six excited states and the ground state of $^{23}$Na for proton energy varying from 
E$_{p}$ = 550 keV to 2 MeV. In a subsequent experiment, G$\ddot{o}$ress \textit{et al.} \cite{gorres1} remeasured the direct capture cross sections elaborately 
from E$_{p}$=550 keV to 1.6 MeV and also deduced the spectroscopic factors of the final bound states from the fit to the capture data. 
The data from recent low energy direct or off-resonance capture measurements by Kelly \textit{et al.} \cite {kelly} at E$_{p}$= 425 keV and by Ferraro, {\it et al.} 
\cite {ferraro} at E$_{p}$=188, 205, 250 and 310 keV beam energies have also been included in the R-matrix analysis. Ferraro {\it et al.} provided S-factor 
data for off-resonance capture to ground state including the contribution of decay of sub-threshold resonance at 8664 keV and the total off-resonance S-factor data. 
The new measurements restricted the R-matrix model prediction for low energy S-factor data for off-resonance capture in $^{22}Ne(p, \gamma)^{23}Na$.

The channel radius, $r_c$, is fixed at $r_c$ = 5.5 fm, a value greater than nuclear radius of $R_N$ = $1.25 \times (A_p^{1/3}+A_T^{1/3}$) 
= 4.75 $fm$ for $^{22}$Ne + $p$ system. Channel radius is not a parameter in R-matrix modelling. A value of $r_c$ = 5.5 fm has been chosen
based on $\chi^{2}$ minimization employing a grid search technique keeping the ANC-s fixed but varying the parameters of the background poles. Search 
has been performed on the total off-resonance S-factor data to choose the radius . 

\begin{figure}
\includegraphics[scale=0.3]{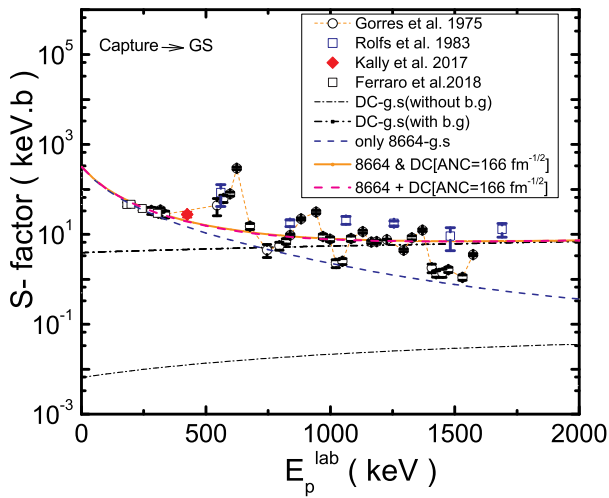}
\caption{\label{fig9} R-matrix fit to the data of direct capture to the ground state of {}$^{23}$Na  }
\end{figure}

\begin{figure}
\includegraphics[scale=0.3]{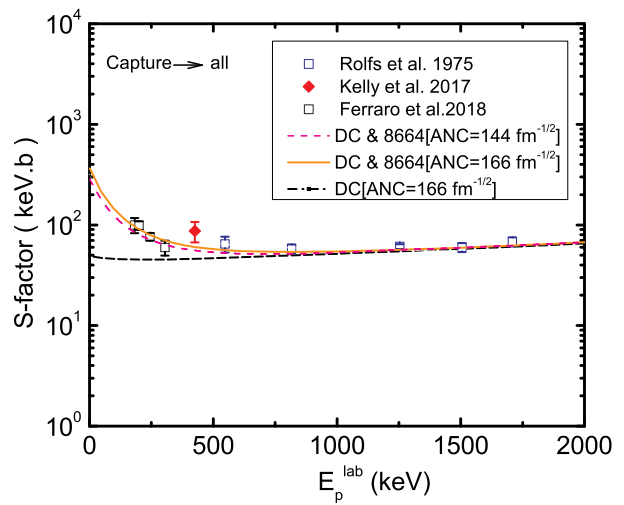}
\caption{\label{fig10} R-matrix fit to the data of total direct capture in {}$^{23}$Na}
\end{figure}

To fit the data for direct capture to individual states and the total off-resonant capture S-factor simultaneously, we consider $M1$ transitions 
to states with $J^\pi$ = 1/2$^+$ ($s$-wave capture), $M1+E2$ transitions to $J^\pi$ = 3/2$^+$ , 5/2$^+$ ($d$-wave capture) and $M1$ transition to 
$J^\pi$ =1/2$^-$ ($p$-wave capture) final bound states. A R-matrix calculation without any background poles or considering only the external contribution 
with the ANC-s listed Tables \ref{tab2} and \ref{tab4} resulted into S(E) curves shown by dashed lines in Fig.~\ref{fig8}.  In Fig. \ref{fig9},
only external capture to the ground state yields the thin black dashed-dotted line  and leads to an extremely low S(0) value. 
Apparently, the external contribution becomes more and more prominent as the excitation energy of the state increases.
To account for the missing cross section, we introduced the high energy background poles in the R-matrix analysis. The poles having spin parity 
1/2$^+$, 1/2$^-$, 3/2$^-$  are included and only the $E1$ decay of the states have been considered. The number of background poles is found to be minimum to 
obtain a simultaneous fit to the data set considered. The poles are placed at an excitation energy of 15 MeV \cite{Kontos}. The proton partial width of the poles are 
fixed at $\Gamma_{p}$ = 5 MeV and it is within the estimated Wigner limit \cite {rolfs book} for particle widths. However, $\Gamma_{\gamma}$ values 
of the background poles are left as free parameters with intial value taken from Weiskoff limit for corresponding gamma transitions.  The fitted background 
pole parameters are shown in Table~\ref{tab6}. The resultant R-matrix fits to the astrophysical S(E) data for direct capture to excited states are shown in 
Fig.~\ref{fig8}. 

In the last panel of Fig.~\ref{fig8}, it is observed that a better fit to DC $\rightarrow$ 8664 keV capture data is obtained for ANC of (166 $\pm$ 4) fm$^{-1/2}$ instead of 144 fm$^{-1/2}$ from transfer calculation. The value and its uncertainty have been obatined from a simultaneous 
best fits, with minimum total $\chi^2$, to DC $\rightarrow$ 8664 keV state, 8664 keV $\rightarrow$ GS and the total S-factor data. In the multiparameter fit,
the background $\Gamma_\gamma$ values are kept free while a grided search is performed over the ANC of the state. The corresponding background pole parameters 
are listed in Table~\ref{tab6}. The condition of simultaneous fitting has reduced the uncertainty in the ANC value.
The enhanced ANC corresponds to spectroscopic factor C$^2$S = 0.43 for the state compared to the value of 
0.32 that yielded ANC = 144 fm$^{-1/2}$ from transfer calculation and corroborates well with the value given 
by Ferraro, {\it et al.} \cite{ferraro}. 

In Fig.~\ref{fig9}, along with the DC $\rightarrow$ GS contribution (black dashed-dotted line), we have shown the contribution from the decay of
8664 keV sub-threshold (-130 keV) resonance (blue dashed line). The state decays to the ground state with a 
branching of (84$\pm$3)$\%$ \cite{firestone} ($\Gamma_\gamma$ = 4.7 eV \cite{gorres}).
The orange solid line in Fig. \ref{fig9} represents the result from R-matrix fit to 
total capture to the ground state of $^{23}$Na. The rise in the low energy S-factor data has been nicely reproduced. No interference effect between the 
two transitions is observed as the summed contribution of individual DC $\rightarrow$ GS and 8664 keV $\rightarrow$ GS (pink dashed line) coincides with 
the solid orange line obtained directly. The total S(E) for off-resonance capture in $^{22}$Ne(p,$\gamma$)$^{23}$Na obtained by summing all the individual 
S(E) functions for transitions to the ground and the excited states is shown in Fig.~\ref{fig10}. Excellent overall fits to the data sets are obtained. 

The total S$_{tot}^{DC}$(0) value for direct capture contribution is 48.8$\pm$9.5 keV.b from the present R-matrix calculation. The uncertainty in the
value includes the contributions from the variation in $r_c$, energy location of background poles and uncertainty values of the ANCs added in quadrature. 
Dominating contribution comes from the uncertainty in the ANC of 8664 keV sub-threshold state. Also a 10$\%$ variation in $r_c$ introduces a variation
of 6.24 keV.b in total direct capture S-factor. The present value is close to S$_{tot}^{DC}$(0) = (50$\pm$12) kev.b reported by Ferraro \cite{ferraro} but 
less than the previously adopted value of 62 keV.b \cite{gorres,angulo}. However, the resultant uncertainty from the present estimation is less.   

\begin{figure}
\includegraphics[scale=0.33]{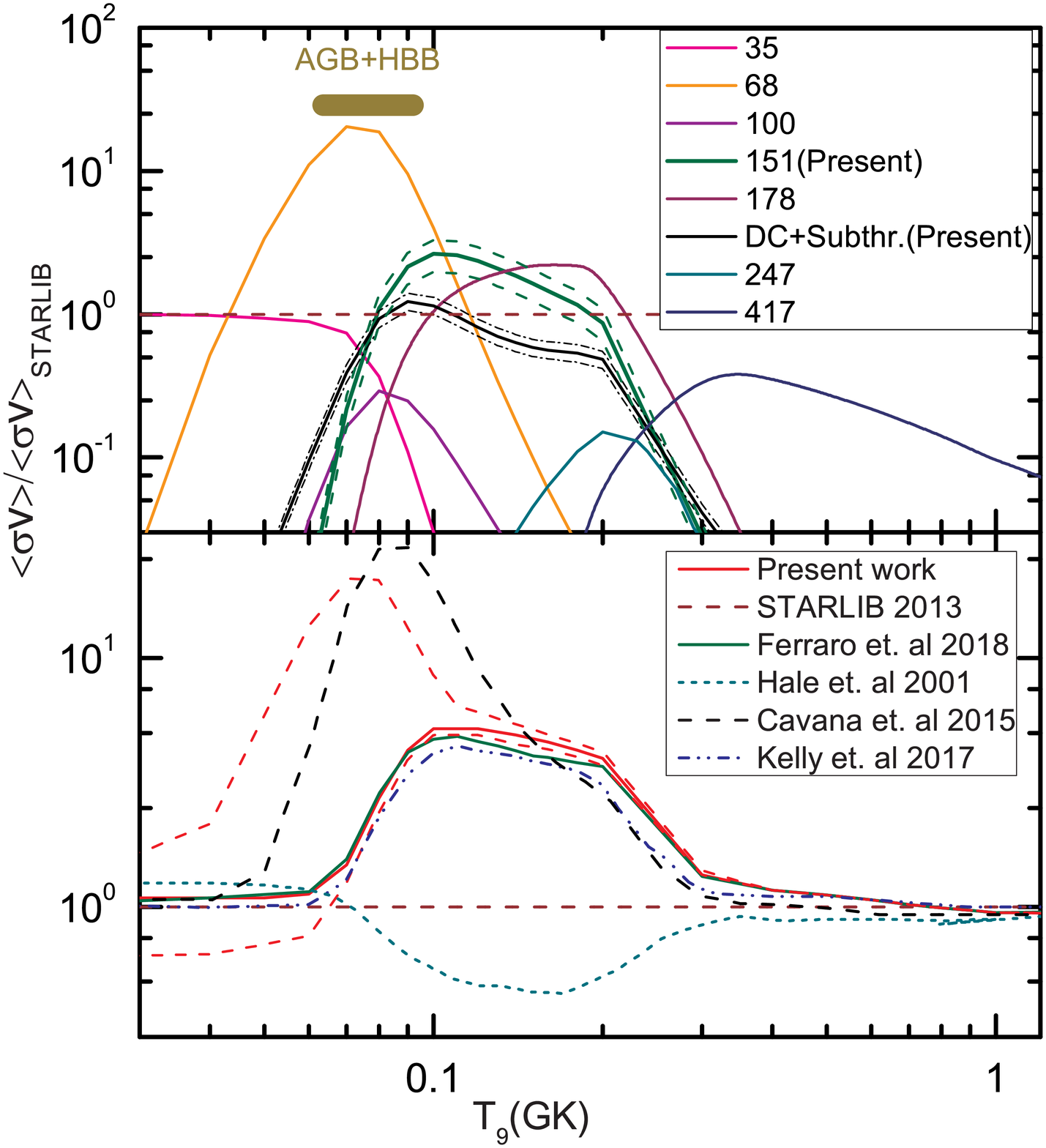}
\caption{\label{fig11} Ratio of reaction rate from present calculation to the STARLIB rate for direct and resonant captures in $^{22}Ne$(p,$\gamma$)$^{23}$Na reaction. 
The solid curve represents the ratio with the total capture rate of the reaction. }
\end{figure}

\begin{table}
\caption{Summary of resonance strengths ($\omega \gamma$) used in reaction rate estimation.}
\label{tab7}
\begin{tabular}{cccccccccccc}    \\ \hline \hline
E$_x$&&& E$_r$&&&&$\omega \gamma$(keV)\footnote{Resonance strengths of states with $E_r$ above 458 keV are taken from STARLIB Compilation \cite{STARLIB}}&&&&$\omega \gamma$(keV) \\ 
(keV)&&&(keV)&&&&Literature&&&&Present \\ \hline
&&&&&&&& \\
8830&&&35&&&&(3.6$\pm$0.2)$\times$10$^{-15}$ \cite{Hale}&&&& \\ 
&&&&&&&& \\
8862&&&68&&&&$\leq$ 6$\times$10$^{-11}$ \cite{ferraro}&&&& \\ 
&&&&&&&& \\
8894&&&100&&&& $\leq$ 7.0$\times$10$^{-11}$ \cite{ferraro}&&&& \\ 
&&&&&&&& \\
8945&&&151&&&& &&&&(2.0$\pm$0.5)$\times$10$^{-7}$ \\ 
&&&&&&&& \\
8944&&&150&&&& &&&&(3.93$\pm$0.9)$\times$10$^{-9}$ \\ 
&&&&&&&& \\
8972&&&178&&&&(2.7$\pm$0.2)$\times$10$^{-6}$ \cite{ferraro}&&&& \\ 
&&&&&&&& \\
9000&&&205.6&&&&$\leq$ 2.8$\times$10$^{-8}$ \cite{cavanna1}&&&&\\ 
&&&&&&&& \\
9042&&&248.4&&&&(9.7$\pm$0.7)$\times$10$^{-6}$ \cite{ferraro}&&&& \\ 
&&&&&&&& \\
9211&&&417&&&&(8.8$\pm$1.02)$\times$10$^{-2}$ \cite{kelly}&&&& \\ 
&&&&&&&& \\  
9252&&&458&&&&0.5 \cite{Meyer}&&&& \\ \hline
&&&&&&&& \\ 
\end{tabular}
\end{table}

\section{Thermonuclear reaction rate of $^{22}Ne (p,\gamma) ^{23}Na$.}

The thermonuclear reaction rate of $^{22}$Ne(p,$\gamma$)$^{23}$Na is controlled by several non interfering low energy narrow resonances and the 
total off-resonance capture reaction. Reaction rate for narrow resonance is calculated using the analytical expression 
\begin{eqnarray}
N_A < \sigma \upsilon> =\sqrt[3]{(\frac {2\pi}{\mu k T}}) \hbar^2 (\omega\gamma)_r exp(-\frac{E_{r}}{kT}). 
\label{eqn2}
\end{eqnarray}
The quantity $\mu$ is the reduced mass, k is Boltzmann$^,s$ constant, $E_{r}$ is resonance energy in the centre of mass frame, and $\omega\gamma$ is the resonance strength with $\omega$
being the statistical spin factor and $\gamma = \frac{\Gamma_p \Gamma_\gamma}{\Gamma}$ where $\Gamma_p$, $\Gamma_\gamma$ and $\Gamma$ are the proton 
partial width, $\gamma$ decay width and the total width, respectively.
Resonance strengths used in the estimation are listed in Table \ref{tab7}. Only the strengths of the doublet states at around $E_r$ = 151 keV have been determined 
in the present work and the corresponding summed contribution is shown by green solid line in the figure. In estimating the reaction rate, all the resonance 
strengths have been divided by the calculated electron screening enhancement factor corresponding to respective excitation energy and tabulated in Table I 
of Ref. \cite{ferraro}. The rates plotted for E$_r$ = 68 and 100 keV 
are calculated with only the experimental upper limits of the respective resonance strengths reported by Ferraro, {\it et al.}. The S$_{tot}$(E), yielded by 
the R-matrix calculation for total DC plus sub-threshold contribution to ground state, is used to get the rate for off-resonant component (black solid line in 
the upper panel). The uncertainty limits of the off-resonant contribution (black dashed dot line) is calculated from the total uncertainty in the off-resonant 
astrophysical S-factor. Individual components are shown in the upper panel of Fig. \ref{fig11}. The non-resonant reaction rates have been determined using the 
code EXP2RATE V2.1 by Thomas Rauscher \cite{rauscher}.  

Total rate, which is the sum of all individual components, is indicated in the figure by a bold red line in the lower panel of Fig. \ref{fig11}. Based on their 
estimation of the upper limits of 68 and 100 keV resonance strengths, Ferraro, {\it et al.} assumed that the role of these resonances in the total rate at relevant 
temperature is insignificant. To compare our total rate in the same temperature window with that of Ferraro, {\it et al.}, Hale, {\it et al.} and 
Cavanna, {\it et al.}, we estimated the total rate without the contributions of 68 and 100 keV resonances. At around
T = 0.1 GK, the present rate is about an order higher than Hale's rate but only slightly higher compared to the 
rates determined by Cavanna as well as Ferraro. For 0.1 GK $\le$ T $\le$ 0.2 GK region, the estimated rate is distinctly higher than both the rates. 
In the T $\le$ 0.1 GK, our rate is similar to the rate of Ref. \cite{ferraro}.
The associated upper and lower uncertainty limits are shown by red dashed lines. While calculating the limits for the resonant capture rate, we took into account 
the uncertainties of the energy locations of states along with respective uncertainties of the resonance strengths. While the upper limits of the strengths of 68 and 100 keV resonances are considered in estimating the upper limit of total rate, for lower limit of the total rate the lower limits of the strengths  
are set to zero \cite{depalo}. Thus the effect of 68 and 100 keV resonances is included in the uncertainty region of the total rate bounded by 
the red dashed lines.

\section{Conclusion} A consistent analysis of direct capture reaction in $^{22}$Ne(p,$\gamma$)$^{23}$Na has been performed within the R-matrix framework. constrained with the asymptotic normalization constants of the bound states of $^{23}$Na obtained from the transfer reaction calculation. Asymptotic normalization constants have been 
extracted from finite DWBA analysis of $^{22}$Ne($^3$He,d)$^{23}$Na transfer data. 

Astrophysical S-factor data for capture to the bound states of $^{23}$Na have been reproduced from the analysis. Contribution of capture through the sub-threshold 
resonance at 8664 keV excitation in the total capture to ground state of $^{23}$Na has been delineated. The observed rise in the ground sate capture data is 
reproduced nicely. Total direct capture S-factor at zero relative energy, $S_{DC}$(0), is found to be 48.8 $\pm$ 9.5 having less uncertainty. 

The total reaction rate obtained as a function of temperature differs from the recent estimations by Ferraro {\it et a.} in the temperature window of 0.1 GK $\le$ T $\le$ 0.2 GK. 
The difference is caused due to slightly higher contribution from direct plus sub-threshold capture to the ground state. However, the present uncertainty in the total rate in this region is relatively higher due to the uncertainty in the resonance strength of the unbound state extracted from transfer angular distribution data. However, in T $\le$ 0.1 GK, the uncertainty in the rate is comparable with the result of Ferraro, {\it et al.}.

\end{document}